# Search for the Neutron Decay n→X+γ, where X is a dark matter particle.

Z. Tang[1], M. Blatnik[2], L. J. Broussard[3], J. H. Choi[4], S. M. Clayton[1],  C. Cude-Woods[1,4], S. Currie[1], D. E. Fellers[1], E. M. Fries[2], P. Geltenbort[5], F. Gonzalez[6], T. M . Ito[1], C.-Y. Liu[6], S. W. T. MacDonald[1],  M. Makela[1], C. L. Morris[1], C. M. O'Shaughnessy[1], R. W. Pattie Jr.[1], B. Plaster[7], D. J. Salvat[8], A. Saunders[1], Z. Wang[1], A. R. Young[1,4], and B. A. Zeck[1,4]

[1]*Los Alamos National Laboratory, Los Alamos, New Mexico 87545, USA*

[2]*Kellogg Radiation Laboratory, California Institute of Technology, Pasadena, CA 91125, USA*

[3]*Oak Ridge National Laboratory, Oak Ridge, Tennessee 37831, USA*

[4]*North Carolina State University, Raleigh, North Carolina 27695, USA*

[5]*Institut Laue-Langevin, Grenoble, France*

[6]*Department of Physics, Indiana University, Bloomington, Indiana 47408, USA*

[7]*University of Kentucky, Lexington, Kentucky 40506, USA*

[8]*University of Washington, Seattle, WA 98195-1560, USA*

***Abstract.*** In a recent paper submitted to Physical Review Letters, Fornal and Grinstein have suggested that the discrepancy between two different methods of neutron lifetime measurements, the beam and bottle methods can be explained by a previously unobserved dark matter decay mode, n→X+γ, where X is a dark matter particle. We have performed a search for this decay mode over the allowed range of energies of the monoenergetic gamma ray for X to be a dark matter particle.  We exclude the possibility of a sufficiently strong branch to explain the lifetime discrepancy with greater than 4 sigma confidence.

There is nearly five standard-deviation disagreement[1,2] between measurements of the rate of neutron decay producing protons measured in cold neutron beam experiements[3-5] ($888.0 \pm 2.0$ s) and free neutron lifetime in bottle experiments  [6-8]  ($878.1 \pm 0.5$ s). The cold neutron beam method measures the number of protons emitted from neutron beta decay in a well-characterized neutron beam, and the bottle experiments measure the number of ultra-cold neutrons (UCN) that remain inside a trap after a certain storage time. A longer lifetime from the beam measurements could point to the existence of possible other decay modes of the neutron where a proton is not produced, which was first pointed out by A. Serebrov, and he suggested the discrepancy could be due to neutrons oscillating into mirror neutrons [9,10]. Recently, Fornal and Grinstein have suggested in Ref. [11] that the neutron lifetime discrepancy can be explained if the neutron decayed into a gamma ray, γ, and a dark matter particle, X. The gamma ray has an allowable energy range of 0.782 to 1.664 MeV, where it is bounded from above by the constraint from the stability of $^9$Be and bounded from below by requiring X to be stable.

Here we report the results of a search for γ arising from UCN decaying inside a nickel phosphorous coated [12], 560 l stainless steel bottle. The bottle is filled with UCN from the Los Alamos UCN facility[13] parasitically during the running of the UCNτ experiment[7], with the source operated in production mode. The gamma-rays are detected in a lead shielded, Compton scattering suppressed 140% high purity germanium (HPGe) detector [Figure 1]. The Compton scattering suppression is achieved by an anti-coincidence with an annular bismuth germinate (BGO) detector surrounding the HPGe detector. A gate valve placed upstream controlled loading of UCN into the bottle. The background γ rates were measured with the UCN in production mode and the gate valve closed.

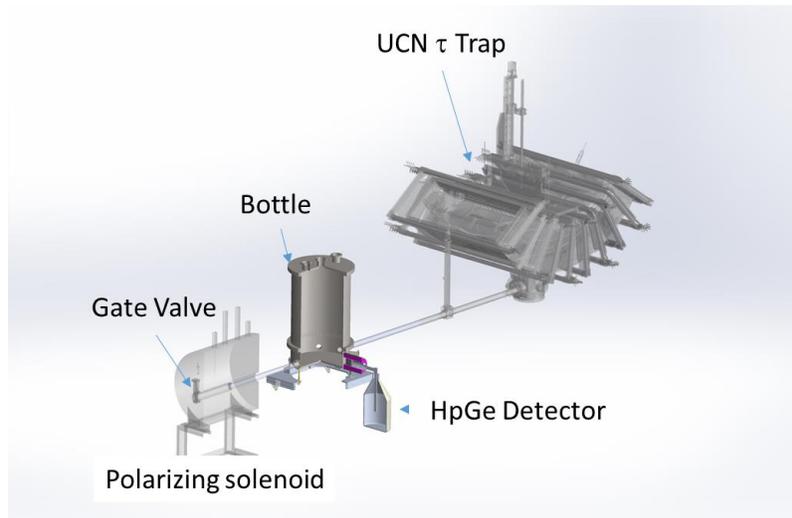

Figure 1: The UCN bottle is installed inline to the existing beamline, and a HPGe detector is placed next to the outer wall of the vessel. The UCN gate valve is located upstream of the polarizing solenoid magnet.

The energy calibration of the HPGe spectrum was obtained from a linear fit to 13 gamma ray lines from sources, natural backgrounds and prominent neutron capture lines on $^{58}$Ni, $^{56}$Fe, and $^{35}$Cl. The UCN induced gamma-ray spectrum was then constructed by subtracting the background spectrum (gate valve closed) from the foreground spectrum (gate valve open). The results of this subtraction are shown in Figure 2. The peaks in the spectrum are dominated by neutron capture lines on the Ni-P surface and in the stainless steel bulk of the storage vessel. The bulk neutron capture is due to UCN upscattering on the surface of the coating. In the region of interest (ROI), we have identified 32 prompt gamma lines from neutron capture on $^{35}$Cl, $^{50}$Cr, $^{52}$Cr, $^{53}$Cr, $^{56}$Fe, $^{58}$Ni, $^{60}$Ni, $^{62}$Ni, $^{63}$Ni, and $^{64}$Ni. There were also 2 lines present from the beta delayed gamma rays from $^{24}$Na, and $^{56}$Mn. Nickel is present both in the coating material and in the bulk stainless steel. Chlorine is used in surface preparation during the nickel phosphorus coating process, and multiple lines outside of the ROI were also identified for Cl. Iron, chromium, and manganese are all components in stainless steel. Sodium-24 is produced in the shielding stack, and it is present in both the foreground and background measurements. However, due to its long half-life (14.96 hours) and the sequential order of the foreground and background measurements, the background subtraction produced a small negative peak.

The strength of each peak inside the ROI was calibrated using the peaks from the same isotope outside of the ROI [14]. A Geant4[15] simulation of the energy dependence of the detector efficiency was used to normalize the peak strength. A Gaussian peak with a 4.2 keV full width at half max and a normalized peak strength was generated for each peak inside the ROI this way, and the sum of all the peaks was then subtracted to obtain the black curve in Figure 2.

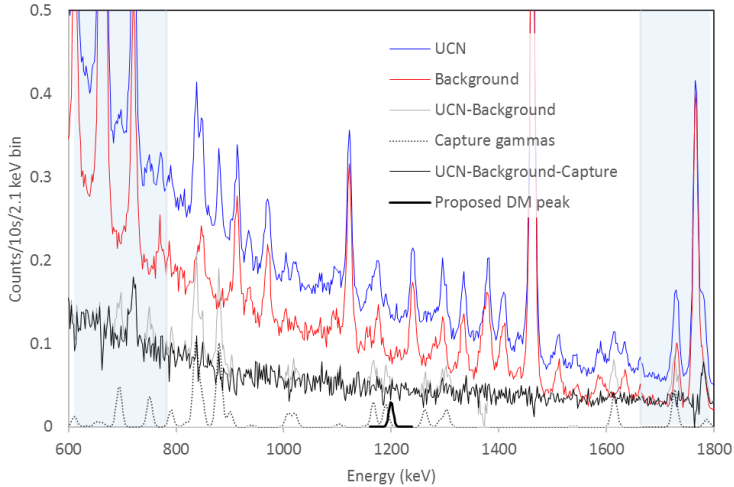

Figure 2: Measured and simulated spectra in the allowed energy region (white background).The blue and red lines show the Compton scattering suppressed spectra for the measurement with UCN and background measurement, respectively.  The dotted line shows the simulated spectra from UCN capture and related gamma rays. The grey and black lines show the net UCN signal and the net signal after capture gamma subtraction, respectively. The peak plotted with a thick black line centered at 1200 keV shows an example of the size of the proposed decay that would be need to explain the anomaly.

To determine the rate of decay into this proposed channel, one needs to know the number of UCN inside the storage volume. The UCN density inside this storage volume was measured using the vanadium activation method[16,17]. A 1.0 cm diameter foil was mounted on the inside of the wall of the vessel, near the detector.  Due to the negative Fermi potential of the $^{51}V$, 84% of UCN that intercept the foil are absorbed and produce $^{52}V$, and a correction is made for neutrons that are upscattered or reflected. Neutron capture on $^{51}V$ produces $^{52}V$, which has a beta decay half-life 3.74 minutes, and a 1434 keV γ is produced along with the beta decay 100% of time. This gamma ray is then detected in the HPGe detector.  The efficiency of the germanium detector was normalized by using a $^{60}Co$ source of known activity (9.3±0.9 kBq) that was placed on top of the $^{51}V$ foil and the rate of 1333 keV γ was measured. This accounted for solid angle and detector efficiency and gamma ray attenuation in the vessel walls. The results were cross calibrated to the measurement by normalizing using upstream $^{10}B/ZnS$ UCN monitor detectors [18].  The average UCN density at beam height in the storage volume for the foreground measurement was $\rho_0$=9.5±1.3  UCN/cm³, where the uncertainty is dominated by the corrections to the $^{51}V$ capture fraction as in ref. [16].

The Ge detector acceptance for gamma rays for each gamma emission position inside the UCN storage vessel was measured by scanning the storage volume with the calibrated $^{60}Co$ source. First, the source was scanned along a line through the center of the detector.  This was fitted with the function $a/(z-z_0)^2$, where z was measured from the cylindrical center of the volume. The constants a and z0 were fitted free parameters.  This determined an effective center of the detector relative to the center of the storage vessel. Next, a 2D counting rate scan was made in two axes, the axis of the cylinder (y) and an axis normal to z and y.  After being normalized to the activity of the source, these data were fitted with a function:

$$R(x, y, z) = \frac{A}{r^2} e^{-\frac{\theta^2}{2\theta_0^2}}$$

(1)

where $\theta_0 = \text{atan}(\frac{\sqrt{x^2+y^2}}{r})$ and $r = \sqrt{x^2 + y^2 + (z - z_0)^2}$.

The acceptance for gamma rays, A, from neutron decay was obtained by integrating this acceptance over the neutron density, assuming a dN/dv ∝ v² distribution, where v is the neutron velocity and N is the neutron density, and accounting for the gravitational distribution of the density[19]:

$$\rho(x, y, z) = \rho_0 \qquad\qquad y < y_{beam}$$
$$= \rho_0 \sqrt{\frac{v_{max}^2 - 2gy}{v_{max}^2}} \quad y_{beam} < y < \frac{v_{max}^2}{2g}$$
$$A = \int_V R(x, y, z) \rho(x, y, z) dV$$

(2)

Where $\rho(x,y,z)$ is the UCN density as a function of position, $v_{max}$ is the maximum UCN velocity, $g$ is the acceleration due to gravity, and the integral is over the volume of the vessel.  A v²dv velocity distribution up to a maximum velocity of 600 cm/s (given by the Fermi potential of upstream stainless steel guides[19]) is used to determine the height dependence of $\rho$.

The volume averaged detector sensitivity is 88 counts/decay/(UCN/cm³). The branching ratio for UCN decay into dark matter needed to explain the difference in the beam and bottle lifetimes is 1.3%. The measured density gives an expected rate of 11 mHz, or 0.11 counts/10 s.

In order to estimate the one sigma uncertainty on detecting a peak we have fitted 100 keV segments of the spectrum in the allowed region with a straight line.  We have then taken the peak detection limit to be $\Delta A = \sqrt{3\sigma_{peak}} RMS$ , where $\sigma_{peak}$ is the HPGe spectrum Gaussian peak width in channels, and $RMS$ is the root mean square of the residual from the fit. These results exclude the presence of a mono-energetic gamma ray from entire allowed region at more than 4 sigma.

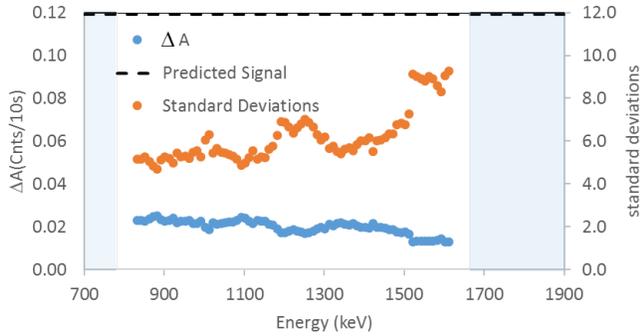

Figure 3: The blue points are one sigma limits on gamma ray peaks in the allowed (white) region for neutron decay into dark matter. The red points are the number of standard deviations away from the expected signal. The presence of a mono-energetic gamma ray is excluded from the entire allowed region by more than 4 sigma.

In summary, we have used the Los Alamos UCN source[13] to search for monoenergetic gamma rays from neutron decay to dark matter, a solution recently proposed to explain the difference between beam and bottle neutron lifetime results.[11] Our measurements exclude this possible explanation[11] with a high degree of confidence.

This work was supported by the Los Alamos Laboratory Directed Research and Development (LDRD) office (No. 20140568DR), the LDRD Program of Oak Ridge National Laboratory, managed by UT-Battelle, LLC (No. 8215), the National Science Foundation (Nos. 130692, 1307426, 161454, 1506459, and 1553861), IU Center for Space Time Symmetries (IUCSS), and DOE Low Energy Nuclear Physics (Nos. DE-FG02-97ER41042, DE-SC0014622,   and DE-AC05-00OR22725). The authors would like to thank the staff of LANSCE for their diligent efforts to develop the diagnostics and new techniques required to provide the proton beam for this experiment.